# Testing the Finch Hypothesis on Green OA Mandate Effectiveness

Y Gargouri [1], V Lariviere [1], Y Gingras [1], T Brody [2], L Carr [2] & S Harnad [1,2]

[1] Université du Québec à Montréal, Canada
[2] University of Southampton, UK

In June 2012, the UK Finch Committee made the following statement:

> *"The [Green OA] policies of neither research funders nor universities themselves have yet had a major effect in ensuring that researchers make their publications accessible in institutional repositories…"*
> [Finch Committee Recommendation, June 2012]

**Testing the Finch Hypothesis**

We have now tested the Finch Hypothesis. Using data from ROARMAP institutional Green OA mandates and data from ROAR on institutional repositories, we found that deposit number and rate is significantly correlated with mandate strength (classified as 1-12): The stronger the mandate, the more the deposits. The strongest mandates generate deposit rates of 70%+ within 2 years of adoption, compared to the un-mandated deposit rate of 20%. The effect is already detectable at the national level, where the UK, which has the largest proportion of Green OA mandates, has a national OA rate of 35%, compared to the global baseline of 25%.

**Conclusion**
The conclusion is that, contrary to the Finch Hypothesis, Green Open Access Mandates *do* have a major effect, and the stronger the mandate, the stronger the effect (the Liege ID/OA mandate, linked to research performance evaluation, being the strongest mandate model). RCUK (as well as all universities, research institutions and research funders worldwide) would be well advised to adopt the strongest Green OA mandates and to integrate institutional and funder mandates.

**Mandate strength:**

    **12**   immediate deposit required + linked to performance evaluation (Liege) (no waiver option)
    **9**   immediate deposit required (no waiver option)
    **6**   6-month delay allowed (no waiver option)
    **3**   12-month delay allowed (no waiver option)
    **3**   rights-retention with waiver option
    **2**   deposit if/when publisher says it's ok
    **1**   no requirement: just request, recommendation or encouragement (policies in ROARMAP that are not classified as mandates)

**Variables:**

- **MAND-STRENGTH**: strength of institution's deposit mandate
- **MAND_AGE**: age of institution's deposit mandate (in months)
- **REPOS_AGE**: age of institution's repository
- **depos_total** : total number of deposits in institutional repository, *normalized by the number the researchers at institution*
- **depos_average** : yearly average number of deposits in institutional repository, *normalized by number of researchers at institution*
- **depos_rate** : rate of deposit (number of days per year with 10-99 deposits)[i], *normalized by the number of researchers at institution*
- **INST_RANK** (12,000 minus Webometrics rank of institution)

We used data from ROARMAP (the *Registry of Open Access Repositories' Mandatory Archiving Policies*) to classify the age (MAND_AGE) (1-103 months) and the strength (MAND-STRENGTH) (1–12) of self-archiving mandates for the 155 institutions that have adopted a mandate to date.

We then used ROAR (the *Registry of Open Access Repositories*) to determine for each registered institutional repository (c. 1,890 repositories to date) its age (REPOS_AGE) and the number (depos_total) rate (depos_rate) and yearly average (depos_average) of its deposits.

Thompson-Reuters database (WoK) was used to normalize the number and the rate of deposits by the number of researchers at each institution that had published at least one article during the last 4 years.

Webometrics university ranking was used to classify institutions according to their rank among 12,000 institutions. INST_RANK (12,000 minus ranking) reversed the rank scale to give the positive correlations a natural interpretation (in which a higher number means a higher rank).

**Negative Binomial Regressions:**

We then used negative binomial regression to test whether mandate MAND-STRENGTH, MAND_AGE, REPOS_AGE and INST_RANK were predictive of the normalized deposit number, average or rate of deposit. These tests were conducted on 5 samples of mandated institutions:

1. All mandated institutions
2. Institutions where MAND_AGE > 2 years
3. Institutions where MAND_AGE > 3 years
4. The top 50% of institutions classified by INST_RANK (INST_RANK < 10,736)
5. The bottom 50% of institutions (10,736 < INST_RANK < 12,000)

**Results:**

The following table shows the significant correlations for each model, tested on each of 5 subsets of institutions. The + sign indicates a significant positive correlation of (Exp(B) < 1.05). The ++ indicates a significant positive correlation of (Exp(B) > 1.05). N is the number of institutions.

| Model | Dependent v. | Independent v. | All Institutions | | Mandate > 2 years | | Mandate > 3 years | | INST_RANK < 10,736 | | INST_RANK > 10,736 | |
|---|---|---|---|---|---|---|---|---|---|---|---|---|
| | | | Exp(B) | N | Exp(B) | N | Exp(B) | N | Exp(B) | N | Exp(B) | N |
| 1 | depos_average | MAND-STRENGTH | ++ | 73 | ++ | 44 |  | 32 |  | 18 | ++ | 55 |
| | | MAND_AGE | + | | | | | | | | + | |
| | | REPOS_AGE | + | | + | | | | + | | + | |
| 2 | depos_rate | MAND-STRENGTH |  | 82 | ++ | 51 |  | 32 |  | 25 |  | 57 |
| | | MAND_AGE | | | | | + | | | | | |
| | | REPOS_AGE | | | + | | | | | | + | |
| 3 | depos_total | MAND-STRENGTH |  | 82 | ++ | 51 | ++ | 32 |  | 25 | + | 57 |
| | | MAND_AGE | + | | | | | | + | | + | |
| | | REPOS_AGE | + | | + | | + | | + | | + | |
| 4 | INST_RANK | MAND-STRENGTH |  | 106 |  | 60 |  | 36 |  | 40 |  | 66 |
| | | MAND_AGE | | | | | | | | | | |
| | | REPOS_AGE | | | | | | | | | | |
| 5 | depos_rate | MAND-STRENGTH |  | 81 | ++ | 50 |  | 31 |  | 25 |  | 56 |
| | | MAND_AGE | | | | | | | | | | |
| | | REPOS_AGE | | | | | | | | | | |
| | | INST_RANK | | | | | | | | | | |
| 6 | depos_total | MAND-STRENGTH |  | 81 | ++ | 50 |  | 31 |  | 25 |  | 56 |
| | | MAND_AGE | + | | | | | | | | | |
| | | REPOS_AGE | + | | ++ | | | | + | | + | |
| | | INST_RANK | | | | | | | | | | |

**Table 1: Significant Positive Correlations for Negative Binomial Regressions:** MAND_AGE, REPOS_AGE, MAND-STRENGTH and INST_RANK tested with normalized number and/or rate of deposit: The stronger the mandate, the higher the deposit number total, average and rate.

**Mandate strength:**

- MAND-STRENGTH is highly and positively correlated with deposit average
- MAND-STRENGTH is highly and positively correlated with deposit rate when mandate age is older than 2 years.
- MAND-STRENGTH is highly and positively correlated with number of deposits when mandate age is older than 3 years (N is as low as 32), as well as when mandate age is older than 2 years (N=51)

**Mandate age:**

- MAND_AGE is positively correlated with deposit number and average
- MAND_AGE is positively correlated with deposit rate when mandate age is older than 3 years (N=32).

**Repository Age:**

- REPOS_AGE is positively correlated with deposit rate and average.
- REPOS_AGE is positively correlated with number of deposits when mandate age is older than 2 years (N = 51), as well as for higher ranked institutions.

**INST_RANK**

There is no significant correlation with INST_RANK, so there is no ranking bias.

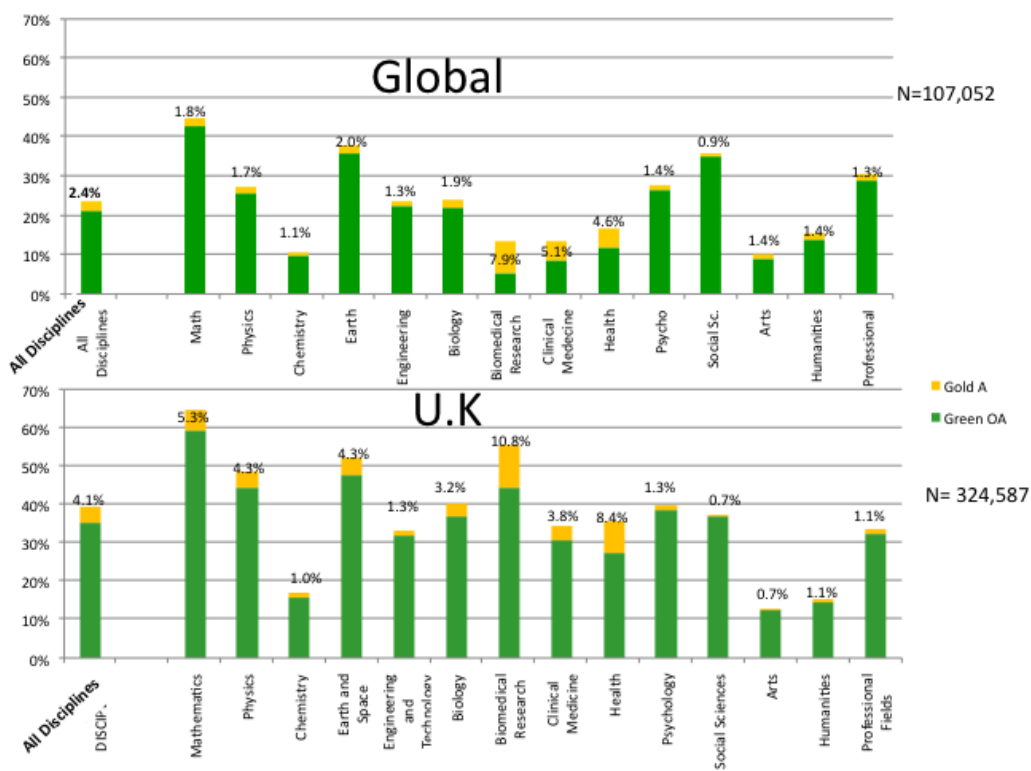

**Comparing percent OA globally and in UK.** UK average OA is close to 40%, 15% higher than the global average of about 25%. The most likely reason is that RCUK as well as about a third of UK universities have mandated Green OA. The UK mandates are not very effective, lacking compliance-verification mechanisms, but they are nevertheless substantially more effective than the global average, which is almost all unmandated OA. UK **Green** OA can be increased to 100% cost-free, by mandating it (effectively). **Gold** OA can only be increased by both mandating it *and* paying publishers extra for it, over and above subscriptions. (Data source: Gargouri et al 2012)

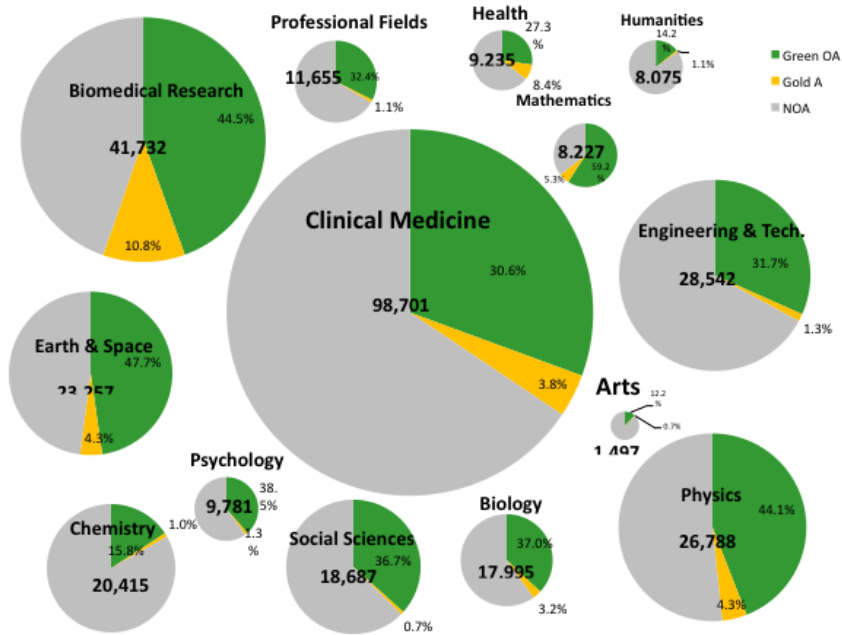

**The Need to Strengthen Green OA Mandates.** As in the rest of the world, most UK OA is **Green**, not **Gold**. UK **Green** OA can be increased to 100% cost-free, by mandating it. **Gold** OA can only be increased by both mandating it and paying publishers extra for it, over and above subscriptions. (Data source: Gargouri 2012)

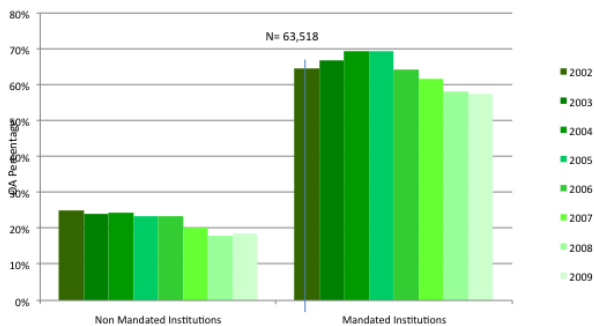

DR2012 Oxford Sept 11 2012

**Effective Mandates**. Effective **Green** OA Mandates generate 70%+ OA. (Data source: Gargouri et al 2010)

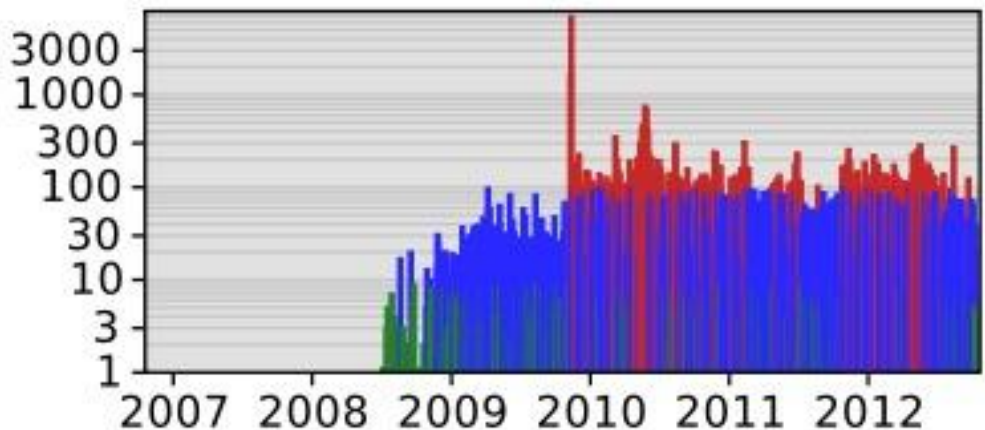

**U. Liège Mandate and Deposit Rate.** U Liège's Rector, Bernard Rentier, reported in 2010 that over the prior year deposits to the U. Liège repository (ORBi) had grown from 10 to 40 thousand publications, 25 thousand of them full-text. This is the highest deposit rate in ROAR's single institutional repository range. Viewed 650 thousand times and downloaded 61 thousand times, the 40 thousand deposits coincided with the first year in which, as a part of U Liège's Open Access Mandate, ORBi was designated as U Liege's sole official means of submitting publications for performance review for academic promotion. (Data source: ROAR)

**References**

How to Integrate University and Funder Open Access Mandates

Integrating Institutional and Funder Open Access Mandates: Belgian Model

Liège ORBi model: Mandatory policy without rights retention but linked to assessment procedures

Optimizing OA Self-Archiving Mandates: What? Where? When? Why? How?

Which Green OA Mandate Is Optimal?

Finch, Dame Janet et al (2012) Accessibility, sustainability, excellence: how to expand access to research publications. *Report of the Working Group on Expanding Access to Published Research Findings*.

---

[i] Deposit rate is calculated as the number of middle-activity days. A low-activity day is 1-9 deposits per day, a middle-activity day is 10-99 deposits per day, and a high-activity day is 100+ deposits per day. ROAR experience has shown that the middle-activity range corresponds to an active single institutional repository's activity level, whereas the high-activity range corresponds to central (multi-institutional) repository activity levels. Central repositories were not included in this study of institutional mandates.